\documentclass[reprint, superscriptaddress, amsmath, amssymb, aps]{revtex4-2}
\usepackage{graphicx,dcolumn,bm,xcolor,mathptmx,etoolbox,blindtext,upgreek,float,nicefrac}
\usepackage[normalem]{ulem}
\usepackage[utf8]{inputenc}
\usepackage[T1]{fontenc}
\definecolor{prlblue}{RGB}{46, 48, 146}
\usepackage[colorlinks=true,citecolor=prlblue,urlcolor=prlblue,linkcolor=prlblue]{hyperref}
\usepackage{silence}
\usepackage[version=3]{mhchem} 
\begin{document}
\title{Implantation of He in Correlated Electron Materials: the Metal-Insulator Transition in SrVO$_3$ }
\author{Carina Jacobson}
\thanks{These authors contributed equally to this work.}
\affiliation{Department of Physics and Astronomy,
Purdue University, West Lafayette, Indiana, USA}
\affiliation{Department of Chemistry,
Indiana University, Bloomington, Indiana, USA}

\author{Md. Rajbanul Akhond}
\thanks{These authors contributed equally to this work.}
\affiliation{Department of Chemistry,
Indiana University, Bloomington, Indiana, USA}

\author{Alexandru B. Georgescu}
\affiliation{Department of Chemistry,
Indiana University, Bloomington, Indiana, USA}

\begin{abstract}
Helium implantation offers a route to tune the electronic properties of correlated oxides through lattice expansion, but its microscopic effects can extend beyond homogeneous strain. Using density functional theory with a Hubbard correction (DFT+$U$), we investigate the structural and electronic response of epitaxially constrained SrVO$_3$ to explicit interstitial He. At a He concentration that produces an out-of-plane expansion of approximately 6\%, structural relaxation also generates octahedral rotations and local distortions of the VO$_6$ coordination environment. These distortions bend V-O-V bonds, lift local orbital degeneracies, and favor an orbitally polarized insulating state. Comparison with uniformly elongated structures reveals an electronic response that cannot be captured by the lattice expansion ratio alone.  Our results identify local structural relaxation as a key component of He-induced electronic reconstruction in SrVO$_3$ and establish the importance of treating implanted He explicitly when modeling strain engineering in correlated materials.
\end{abstract}
\maketitle

\section{Introduction}

The electronic phases of correlated transition-metal oxides are sensitive to the coupling between electronic degrees of freedom and crystal structure. Changes in bond lengths, bond angles, and local coordination modify electron hopping and crystal-field splittings, changing the balance between kinetic energy and electron correlations. These structural changes can stabilize distinct orbital and magnetic states and drive metal-insulator transitions (MITs). Understanding which crystal degrees of freedom control these transitions is key to identifying their microscopic mechanisms and for engineering electronic properties\citep{Georgescu2022}.

\begin{figure*}[t]
\centering
\includegraphics[width=0.72\textwidth]{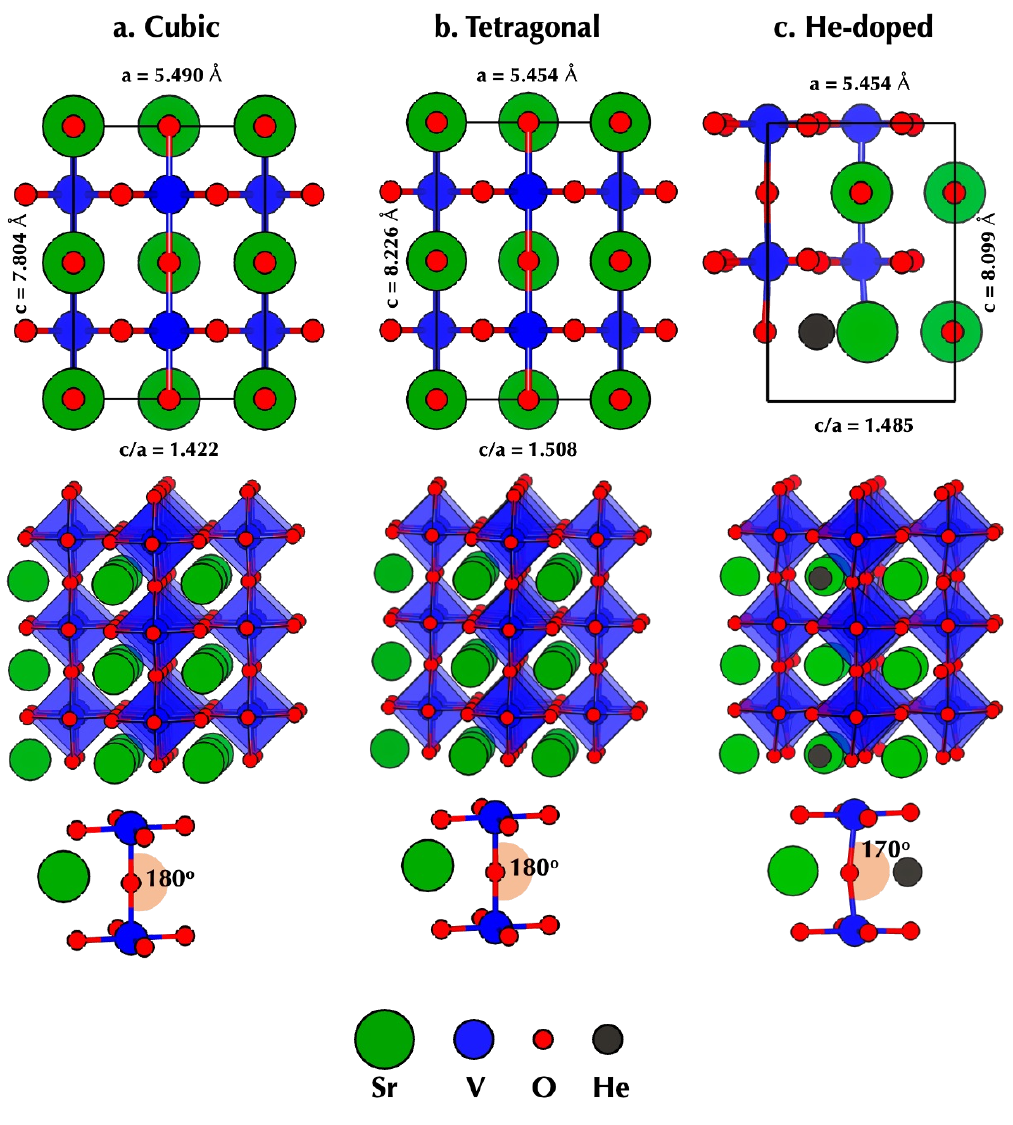}
\caption{Structural comparison of (a) the cubic reference, (b) uniformly elongated He-free SVO, and (c) relaxed He-containing SVO. The upper and middle rows show atomic arrangements and oxygen coordination polyhedra; the lower row highlights representative V-O-V bond angles. The He-free structures retain straight bonds, whereas explicit He incorporation produces local displacements and bond bending to approximately $170^\circ$. Sr, V, O, and He are shown in green, blue, red, and gray, respectively. The displayed lattice dimensions refer to the supercell. For the $\sqrt{2}\times\sqrt{2}\times2$ cell, the pseudocubic ratio is $c_{\mathrm{pc}}/a_{\mathrm{pc}}=c_{\mathrm{sc}}/(\sqrt{2}a_{\mathrm{sc}})$, giving approximately 1.0665 and 1.050 for panels (b) and (c), respectively.}
\label{fig:structures}
\end{figure*}

Epitaxial strain provides a well-established means of controlling the structure of oxide thin films, but the accessible strain states are constrained by the available substrates and the coupling between in-plane and out-of-plane lattice parameters \citep{Herklotz2016b,Herklotz2025a}: the out-of-plane lattice constant relaxes to accomodate the in-plane lattice strain. Ion implantation offers an additional, post-synthesis means of tuning the crystal structure. In epitaxially constrained films, He implantation can continuously increase the out-of-plane lattice parameter while the in-plane parameters remains relatively locked to the substrate. This approach has enabled control of octahedral rotations, magnetic behavior, and optical properties across several oxide systems\citep{Herklotz2016b,Brahlek2023,Herklotz2016a}. Reversibility through annealing has also been demonstrated in selected materials \citep{Brahlek2023}. Because He is chemically inert, its effects are often interpreted primarily in terms of the strain generated by implantation. Nevertheless, an interstitial He atom occupies space within the crystal, and accommodating it can require local atomic displacements as well as a change in the lattice parameters. The electronic consequences of these local distortions need not be captured by a homogeneous-strain description. In certain cases, as shown by two of the authors of this work, He implantation may even lead to defect states that are responsible for electronic states observed in the resulting material, for example in the appearance of magnetism in He-doped PdCoO$_2$, which is explained by the formation of magnetic Co Frenkel pairs \citep{kim2026}.

Studies of other oxides demonstrate that He implantation can modify
electronic properties through several distinct microscopic routes.
Reversible lattice expansion in $La_{0.7}Sr_{0.3}MnO_3$ enables control
of magnetic and metal-insulator transition temperatures
\citep{Guo2015}, while strain doping provides access to tunable
magnetic anisotropy\citep{Herklotz2018Anisotropy} and
post-synthesis control of the anomalous Hall response in
\ce{SrRuO3}\citep{Skoropata2021}. In \ce{BiFeO3}, He implantation
modifies optical spectra through structural changes
\citep{Herklotz2019Optical} and enables spatial patterning of
enhanced tetragonality and structural phase coexistence
\citep{Toulouse2021}. Irradiation can also act through disorder
and defect formation: it drives localization in rare-earth
nickelates\citep{Wang2019Disorder}, selectively modifies transport
in the monoclinic and rutile phases of \ce{VO2}
\citep{Gurrola2024}.
The importance of separating structural and electronic effects
is further illustrated by bulk and layer-confined nickelates,
where electronic confinement and the elastic response of adjacent
layers make distinct contributions to the metal-insulator
transition\citep{Georgescu2019}. These findings motivate an
explicit examination of how implanted He changes local
coordination, beyond its effect on the average lattice parameters.

The sensitivity of oxide properties to their local environment provides several routes for controlling collective electronic behavior. For example, engineering the periodicity of La$_2$NiMnO$_6$/Nd$_2$NiMnO$_6$ superlattices tunes their magnetic transition temperatures and enhances rare-earth exchange interactions \citep{Spring2025}. Post-synthesis helium implantation offers complementary control: it can activate ferroelectricity in HfO$_2$-based films \citep{Kang2022}, while helium-induced defects and lattice distortions in SrRuO$_3$ have been associated with a reported topological Hall response \citep{Wang2020Hall}. These examples motivate examining how the local structural changes accompanying helium implantation influence electronic properties beyond the effects captured by average lattice parameters alone.

In perovskite oxides, octahedral rotations and distortions of the local coordination environment couple lattice structure to orbital and magnetic degrees of freedom\citep{Bhattacharya2014}. Ion implantation provides a route to manipulate these structural degrees of freedom, extending its potential beyond control of the average lattice parameters to orbital engineering \citep{Herklotz2025a}.

The metallic perovskite \ce{SrVO3} (SVO) provides a useful setting in which to examine this distinction. Its nominal V $d^1$ configuration places one electron in the $t_{2g}$ manifold, making its electronic state sensitive to changes in both orbital degeneracy and intersite hopping. Wang \textit{et al.} reported an MIT in He-irradiated SVO films grown on \ce{SrTiO3}, accompanied by an increase in the out-of-plane lattice parameter at fixed in-plane lattice constants\citep{Wang2019}. Their transport measurements supported a substantial role for electron-electron interactions, while their density functional theory plus dynamical mean-field theory (DFT+DMFT) calculations found enhanced correlations under tetragonal expansion without obtaining an insulating solution. More recently, Pickem \textit{et al.} reported that insulating solutions can be stabilized within a tetragonal DFT+DMFT model, emphasizing the role of crystal-field splitting and identifying a region of metallic and insulating coexistence \citep{pickem2026}. Their structural model excludes bond distortions such as octahedral rotations. 

These studies motivate a complementary microscopic question: what local structural changes accompany the explicit incorporation of He, and how do those changes affect the electronic state?

This question is particularly relevant because He implantation and lattice expansion are coupled. The concentration of implanted He that produces a given out-of-plane expansion also determines the local environment that the host lattice must accommodate. If substantial local distortions develop at that concentration, they form part of the structural response responsible for the expansion. In SVO, bending V-O-V bonds can modify hopping between neighboring V sites, while distortions of the VO$_6$ coordination environment can lift local orbital degeneracies. A description based only on the ratio $c/a$ therefore leaves open the electronic consequences of the internal atomic displacements generated by implantation.

Here, we use plane-wave density functional theory with a Hubbard correction (DFT+$U$) to investigate explicit interstitial He in epitaxially constrained SVO and compare its effects with those of homogeneous tetragonal elongation. We find that He incorporation produces coupled out-of-plane expansion and local symmetry lowering, including octahedral rotations and Jahn-Teller-like distortions. At a He implantation that yields $c/a \approx 1.06$, these local distortions accompany the macroscopic expansion and favor an orbitally polarized insulating state. Our calculations identify changes in the local V-O coordination as a consequential component of the electronic response to He implantation. The results establish the importance of considering the microscopic accommodation of implanted He alongside the average lattice deformation when interpreting and designing electronic phase control in correlated oxides.

\section{Results}

To distinguish the effects of homogeneous expansion, local structural relaxation, and the direct presence of He, we compare four configurations: (i) cubic \ce{SrVO3}, (ii) He-free SVO with an imposed tetragonal elongation, (iii) SVO containing explicit interstitial He, with the internal coordinates and out-of-plane lattice parameter relaxed, and (iv) the relaxed He-containing structure with He removed and the host atomic positions and lattice vectors held fixed. The cubic and uniformly elongated structures remain metallic, whereas both the relaxed He-containing structure and its frozen, He-free counterpart exhibit a band gap. Unless otherwise stated, the results discussed below correspond to $U=4$~eV.

\begin{figure*}[t]
\centering
\includegraphics[width=0.67\textwidth]{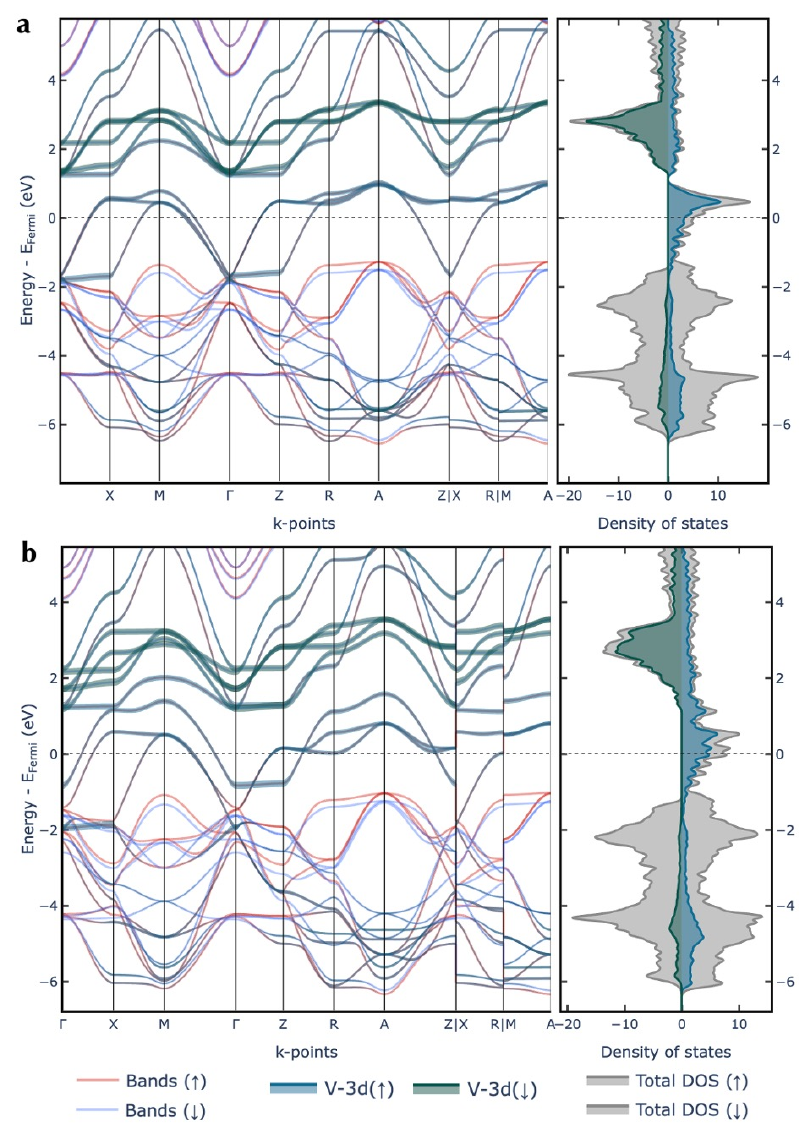}
\caption{Spin-resolved electronic band structures and densities of states for (a) cubic SVO and (b) uniformly elongated He-free SVO. Red and blue lines distinguish the spin channels, and the projected weights highlight V $3d$ character. The accompanying densities of states show the total and V $3d$ contributions, with the spin channels plotted on opposite sides of the vertical axis. Energies are referenced to the Fermi level, indicated by the horizontal dashed lines. Both structures retain bands crossing the Fermi level and remain metallic.}
\label{fig:reference_bands}
\end{figure*}

The cubic reference structure remains metallic over the investigated Hubbard parameters, $U=2$, 3, and 4~eV. Its straight V-O-V bonds provide the reference geometry for assessing the effects of expansion and local distortion. A ferromagnetic solution can be stabilized within DFT+$U$, although bulk SVO is experimentally paramagnetic. We also considered A-, C-, and G-type antiferromagnetic configurations to examine the dependence of the calculated electronic state on magnetic order.

We first examine homogeneous tetragonal elongation without introducing He. The structure shown has supercell dimensions $a_{\mathrm{sc}}=5.454$~\AA{} and $c_{\mathrm{sc}}=8.226$~\AA{}, corresponding to $c_{\mathrm{pc}}/a_{\mathrm{pc}}\approx1.0665$. The V-O-V bonds remain straight, and the band structure and density of states retain states crossing the Fermi level. Thus, an elongation of this magnitude does not open a gap in the He-free reference calculation.

\begin{figure*}[t]
\centering
\includegraphics[width=0.8\textwidth]{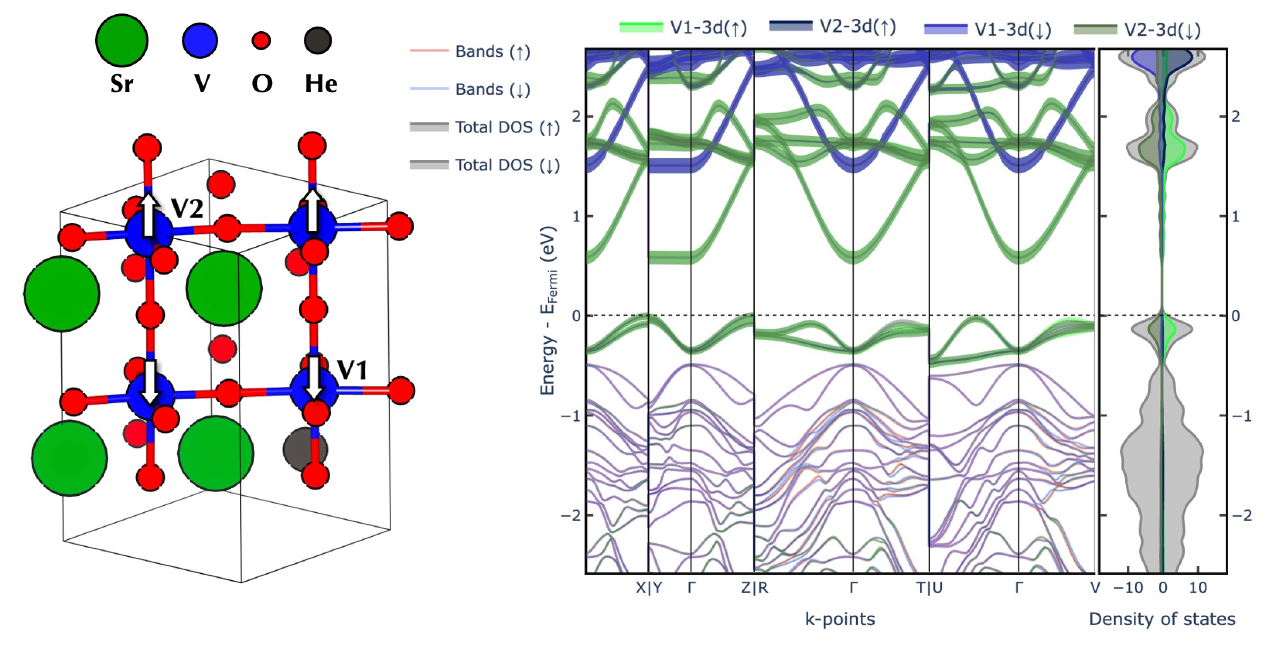}
\caption{Electronic structure of He-containing SVO and comparison of orbital-resolved densities of states. The He-containing structure with A-type antiferromagnetic order, together with its band structure and density of states. Arrows indicate the spin arrangement, and the projected bands distinguish the V1 and V2 contributions in each spin channel. A gap separates the occupied and unoccupied states. }
\label{fig:he_bands_orbitals}
\end{figure*}

We next introduce one He atom per four SVO formula units and relax the internal coordinates and out-of-plane lattice parameter while fixing the in-plane lattice parameters. For the initial interstitial positions examined, He relaxes to the center of a square formed by neighboring V sites. The relaxed He-containing structure has $a_{\mathrm{sc}}=5.454$~\AA{} and $c_{\mathrm{sc}}=8.099$~\AA{}, giving $c_{\mathrm{pc}}/a_{\mathrm{pc}}\approx1.050$. Among the magnetic configurations examined, the lowest-energy He-containing structure has A-type antiferromagnetic order.

Accommodating He also produces substantial internal atomic displacements. The structural comparison shows octahedral rotations and distortions of the local VO$_6$ coordination environment, with representative V-O-V angles reduced from $180^\circ$ to approximately $170^\circ$. These local changes accompany the out-of-plane expansion and are not specified by the lattice-parameter ratio alone.

The relaxed He-containing structure is insulating, with a calculated band gap of approximately $0.512$~eV. Its orbital-resolved density of states shows a redistribution of V $3d$ spectral weight consistent with orbital polarization accompanying the local symmetry lowering. Bending the V-O-V bonds can modify intersite hopping, while distortions of the coordination environment can lift local orbital degeneracies. Notably, the He-containing structure is insulating despite having a smaller pseudocubic lattice-parameter ratio than the metallic, uniformly elongated reference. The difference in electronic behavior therefore cannot be attributed simply to a larger homogeneous elongation in the He-containing calculation.

To separate the electronic effect of the distorted host geometry from the direct presence of He, we remove He from the relaxed structure and recalculate the electronic state with the host atomic positions and lattice vectors frozen. The resulting He-free configuration retains a band gap. This demonstrates that the distorted host geometry can sustain an insulating state within DFT+$U$ without the direct electronic contribution of He. The frozen structure serves as a diagnostic configuration; it does not imply that the distortion would persist after He removal if further structural relaxation were allowed.

Together, these comparisons identify local structural relaxation as a consequential component of the electronic response to He implantation. Homogeneously elongated SVO remains metallic in the calculation shown, whereas the locally distorted host supports an insulating state both with and without He present. Explicit He incorporation thus produces coupled macroscopic expansion and local symmetry lowering whose electronic consequences extend beyond those captured by homogeneous tetragonal strain.

\section{Methods}

We performed spin-polarized plane-wave density functional theory calculations using the \textsc{Quantum ESPRESSO} package\citep{giannozzi2017}, the Perdew-Burke-Ernzerhof (PBE) exchange-correlation functional\citep{Perdew1996}, and ultrasoft pseudopotentials\citep{Vanderbilt1990}. Electron correlations in the V $3d$ manifold were treated using the simplified rotationally invariant DFT+$U$ approach of Dudarev \textit{et al.}\citep{Dudarev1998}, with atomic Hubbard projectors. We considered $U=2$, 3, and 4~eV, focusing on $U=4$~eV unless otherwise stated.

The kinetic-energy cutoffs for the wavefunctions and charge density were 612.3 and 6122.6~eV, respectively. Brillouin-zone sampling used a uniform $4\times4\times4$ $k$-point mesh, with Marzari-Vanderbilt cold smearing of width 0.02~Ry\citep{Marzari1999}. To model epitaxial constraint, we fixed the in-plane lattice vectors while allowing the out-of-plane lattice parameter and internal atomic coordinates to relax in the He-containing structures. For the frozen-structure control, we removed He from the relaxed structure and recalculated the electronic state without changing the host atomic positions or lattice vectors.

\section{Discussion }

Our results show that the electronic response of SVO to He implantation depends on both the average lattice deformation and the local structural changes required to accommodate interstitial He. The relaxed He-containing structure is insulating even though its pseudocubic $c/a$ ratio is smaller than that of the metallic, uniformly elongated reference. Moreover, the gap persists when He is removed while the host geometry is held fixed. These comparisons identify the distorted host structure as sufficient to sustain an insulating state within our DFT+$U$ calculations, without requiring the direct electronic contribution of He.

He implantation, lattice expansion, and local distortion are coupled aspects of the implantation process. At the modeled concentration, accommodating He bends V-O-V bonds and lowers the local symmetry of the VO$_6$ coordination environment. These changes can modify intersite hopping and orbital splittings, providing electronic control beyond that described by the lattice-parameter ratio alone. Homogeneous strain calculations can therefore capture an important component of the response while omitting the local structural effects of the atoms responsible for the expansion.

This interpretation complements the recent report of insulating solutions in a tetragonal DFT+DMFT model\citep{pickem2026}. The existence of such solutions does not establish that homogeneous expansion fully describes the implanted material. Conversely, our ordered DFT+$U$ calculations do not determine the finite-temperature phase diagram or establish a unique mechanism for the experimental transition. They demonstrate that explicit He incorporation generates local distortions with consequential electronic effects.

More broadly, He implantation offers a means of controlling local coordination as well as average lattice parameters in correlated materials. In hosts with multiple inequivalent interstitial sites or preexisting structural distortions, the location and arrangement of implanted He may influence which lattice modes are activated and how they couple to the electronic state. Accounting for this microscopic structural response will be important when using implantation to engineer electronic phases.


\section*{Acknowledgments}
This research was supported by Indiana University startup funds, and in part by the Lilly Endowment, Inc., through its support for the Indiana University Pervasive Technology Institute.

\bibliography{SVO}

\begin{thebibliography}{25}%
\makeatletter
\providecommand \@ifxundefined [1]{%
 \@ifx{#1\undefined}
}%
\providecommand \@ifnum [1]{%
 \ifnum #1\expandafter \@firstoftwo
 \else \expandafter \@secondoftwo
 \fi
}%
\providecommand \@ifx [1]{%
 \ifx #1\expandafter \@firstoftwo
 \else \expandafter \@secondoftwo
 \fi
}%
\providecommand \natexlab [1]{#1}%
\providecommand \enquote  [1]{``#1''}%
\providecommand \bibnamefont  [1]{#1}%
\providecommand \bibfnamefont [1]{#1}%
\providecommand \citenamefont [1]{#1}%
\providecommand \href@noop [0]{\@secondoftwo}%
\providecommand \href [0]{\begingroup \@sanitize@url \@href}%
\providecommand \@href[1]{\@@startlink{#1}\@@href}%
\providecommand \@@href[1]{\endgroup#1\@@endlink}%
\providecommand \@sanitize@url [0]{\catcode `\\12\catcode `\$12\catcode `\&12\catcode `\#12\catcode `\^12\catcode `\_12\catcode `\%12\relax}%
\providecommand \@@startlink[1]{}%
\providecommand \@@endlink[0]{}%
\providecommand \url  [0]{\begingroup\@sanitize@url \@url }%
\providecommand \@url [1]{\endgroup\@href {#1}{\urlprefix }}%
\providecommand \urlprefix  [0]{URL }%
\providecommand \Eprint [0]{\href }%
\providecommand \doibase [0]{https://doi.org/}%
\providecommand \selectlanguage [0]{\@gobble}%
\providecommand \bibinfo  [0]{\@secondoftwo}%
\providecommand \bibfield  [0]{\@secondoftwo}%
\providecommand \translation [1]{[#1]}%
\providecommand \BibitemOpen [0]{}%
\providecommand \bibitemStop [0]{}%
\providecommand \bibitemNoStop [0]{.\EOS\space}%
\providecommand \EOS [0]{\spacefactor3000\relax}%
\providecommand \BibitemShut  [1]{\csname bibitem#1\endcsname}%
\let\auto@bib@innerbib\@empty
\bibitem [{\citenamefont {Georgescu}\ and\ \citenamefont {Millis}(2022)}]{Georgescu2022}%
  \BibitemOpen
  \bibfield  {author} {\bibinfo {author} {\bibfnamefont {A.~B.}\ \bibnamefont {Georgescu}}\ and\ \bibinfo {author} {\bibfnamefont {A.~J.}\ \bibnamefont {Millis}},\ }\bibfield  {title} {\bibinfo {title} {Quantifying the role of the lattice in metal--insulator phase transitions},\ }\href {https://doi.org/10.1038/s42005-022-00909-z} {\bibfield  {journal} {\bibinfo  {journal} {Communications Physics}\ }\textbf {\bibinfo {volume} {5}},\ \bibinfo {pages} {135} (\bibinfo {year} {2022})}\BibitemShut {NoStop}%
\bibitem [{\citenamefont {Herklotz}\ \emph {et~al.}(2016{\natexlab{a}})\citenamefont {Herklotz}, \citenamefont {Wong}, \citenamefont {Meyer}, \citenamefont {Biegalski}, \citenamefont {Lee},\ and\ \citenamefont {Ward}}]{Herklotz2016b}%
  \BibitemOpen
  \bibfield  {author} {\bibinfo {author} {\bibfnamefont {A.}~\bibnamefont {Herklotz}}, \bibinfo {author} {\bibfnamefont {A.~T.}\ \bibnamefont {Wong}}, \bibinfo {author} {\bibfnamefont {T.}~\bibnamefont {Meyer}}, \bibinfo {author} {\bibfnamefont {M.~D.}\ \bibnamefont {Biegalski}}, \bibinfo {author} {\bibfnamefont {H.~N.}\ \bibnamefont {Lee}},\ and\ \bibinfo {author} {\bibfnamefont {T.~Z.}\ \bibnamefont {Ward}},\ }\bibfield  {title} {\bibinfo {title} {Controlling octahedral rotations in a perovskite via strain doping},\ }\href {https://doi.org/10.1038/srep26491} {\bibfield  {journal} {\bibinfo  {journal} {Scientific Reports}\ }\textbf {\bibinfo {volume} {6}},\ \bibinfo {pages} {26491} (\bibinfo {year} {2016}{\natexlab{a}})}\BibitemShut {NoStop}%
\bibitem [{\citenamefont {Herklotz}\ \emph {et~al.}(2025)\citenamefont {Herklotz}, \citenamefont {Petrie},\ and\ \citenamefont {Ward}}]{Herklotz2025a}%
  \BibitemOpen
  \bibfield  {author} {\bibinfo {author} {\bibfnamefont {A.}~\bibnamefont {Herklotz}}, \bibinfo {author} {\bibfnamefont {J.~R.}\ \bibnamefont {Petrie}},\ and\ \bibinfo {author} {\bibfnamefont {T.~Z.}\ \bibnamefont {Ward}},\ }\bibfield  {title} {\bibinfo {title} {Beyond epitaxy: Ion implantation as a tool for orbital engineering},\ }\href {https://doi.org/10.1021/acsaelm.5c00815} {\bibfield  {journal} {\bibinfo  {journal} {ACS Applied Electronic Materials}\ }\textbf {\bibinfo {volume} {7}},\ \bibinfo {pages} {7580} (\bibinfo {year} {2025})},\ \Eprint {https://arxiv.org/abs/https://doi.org/10.1021/acsaelm.5c00815} {https://doi.org/10.1021/acsaelm.5c00815} \BibitemShut {NoStop}%
\bibitem [{\citenamefont {Brahlek}\ \emph {et~al.}(2023)\citenamefont {Brahlek}, \citenamefont {Mazza}, \citenamefont {Annaberdiyev}, \citenamefont {Chilcote}, \citenamefont {Rimal}, \citenamefont {Halász}, \citenamefont {Pham}, \citenamefont {Pai}, \citenamefont {Krogel}, \citenamefont {Lapano}, \citenamefont {Lawrie}, \citenamefont {Eres}, \citenamefont {McChesney}, \citenamefont {Prokscha}, \citenamefont {Suter}, \citenamefont {Oh}, \citenamefont {Freeland}, \citenamefont {Cao}, \citenamefont {Gardner}, \citenamefont {Salman}, \citenamefont {Moore}, \citenamefont {Ganesh},\ and\ \citenamefont {Ward}}]{Brahlek2023}%
  \BibitemOpen
  \bibfield  {author} {\bibinfo {author} {\bibfnamefont {M.}~\bibnamefont {Brahlek}}, \bibinfo {author} {\bibfnamefont {A.~R.}\ \bibnamefont {Mazza}}, \bibinfo {author} {\bibfnamefont {A.}~\bibnamefont {Annaberdiyev}}, \bibinfo {author} {\bibfnamefont {M.}~\bibnamefont {Chilcote}}, \bibinfo {author} {\bibfnamefont {G.}~\bibnamefont {Rimal}}, \bibinfo {author} {\bibfnamefont {G.~B.}\ \bibnamefont {Halász}}, \bibinfo {author} {\bibfnamefont {A.}~\bibnamefont {Pham}}, \bibinfo {author} {\bibfnamefont {Y.-Y.}\ \bibnamefont {Pai}}, \bibinfo {author} {\bibfnamefont {J.~T.}\ \bibnamefont {Krogel}}, \bibinfo {author} {\bibfnamefont {J.}~\bibnamefont {Lapano}}, \bibinfo {author} {\bibfnamefont {B.~J.}\ \bibnamefont {Lawrie}}, \bibinfo {author} {\bibfnamefont {G.}~\bibnamefont {Eres}}, \bibinfo {author} {\bibfnamefont {J.}~\bibnamefont {McChesney}}, \bibinfo {author} {\bibfnamefont {T.}~\bibnamefont {Prokscha}}, \bibinfo {author} {\bibfnamefont {A.}~\bibnamefont {Suter}}, \bibinfo {author} {\bibfnamefont
  {S.}~\bibnamefont {Oh}}, \bibinfo {author} {\bibfnamefont {J.~W.}\ \bibnamefont {Freeland}}, \bibinfo {author} {\bibfnamefont {Y.}~\bibnamefont {Cao}}, \bibinfo {author} {\bibfnamefont {J.~S.}\ \bibnamefont {Gardner}}, \bibinfo {author} {\bibfnamefont {Z.}~\bibnamefont {Salman}}, \bibinfo {author} {\bibfnamefont {R.~G.}\ \bibnamefont {Moore}}, \bibinfo {author} {\bibfnamefont {P.}~\bibnamefont {Ganesh}},\ and\ \bibinfo {author} {\bibfnamefont {T.~Z.}\ \bibnamefont {Ward}},\ }\bibfield  {title} {\bibinfo {title} {Emergent magnetism with continuous control in the ultrahigh-conductivity layered oxide pdcoo2},\ }\href {https://doi.org/10.1021/acs.nanolett.3c01065} {\bibfield  {journal} {\bibinfo  {journal} {Nano Letters}\ }\textbf {\bibinfo {volume} {23}},\ \bibinfo {pages} {7279} (\bibinfo {year} {2023})}\BibitemShut {NoStop}%
\bibitem [{\citenamefont {Herklotz}\ \emph {et~al.}(2016{\natexlab{b}})\citenamefont {Herklotz}, \citenamefont {Rus},\ and\ \citenamefont {Ward}}]{Herklotz2016a}%
  \BibitemOpen
  \bibfield  {author} {\bibinfo {author} {\bibfnamefont {A.}~\bibnamefont {Herklotz}}, \bibinfo {author} {\bibfnamefont {S.~F.}\ \bibnamefont {Rus}},\ and\ \bibinfo {author} {\bibfnamefont {T.~Z.}\ \bibnamefont {Ward}},\ }\bibfield  {title} {\bibinfo {title} {Continuously controlled optical band gap in oxide semiconductor thin films},\ }\href {https://doi.org/10.1021/acs.nanolett.5b04815} {\bibfield  {journal} {\bibinfo  {journal} {Nano Letters}\ }\textbf {\bibinfo {volume} {16}},\ \bibinfo {pages} {1782} (\bibinfo {year} {2016}{\natexlab{b}})}\BibitemShut {NoStop}%
\bibitem [{\citenamefont {Kim}\ \emph {et~al.}(2026)\citenamefont {Kim}, \citenamefont {Samanta}, \citenamefont {Jacobson}, \citenamefont {Chen}, \citenamefont {Lasseter}, \citenamefont {Mallick}, \citenamefont {Cook}, \citenamefont {Yuan}, \citenamefont {Oh}, \citenamefont {Eres}, \citenamefont {Moore}, \citenamefont {Prokscha}, \citenamefont {Suter}, \citenamefont {Randolph}, \citenamefont {Gardner}, \citenamefont {Rack}, \citenamefont {Ward}, \citenamefont {Salman}, \citenamefont {Georgescu},\ and\ \citenamefont {Brahlek}}]{kim2026}%
  \BibitemOpen
  \bibfield  {author} {\bibinfo {author} {\bibfnamefont {S.}~\bibnamefont {Kim}}, \bibinfo {author} {\bibfnamefont {B.}~\bibnamefont {Samanta}}, \bibinfo {author} {\bibfnamefont {C.}~\bibnamefont {Jacobson}}, \bibinfo {author} {\bibfnamefont {A.-H.}\ \bibnamefont {Chen}}, \bibinfo {author} {\bibfnamefont {J.~C.}\ \bibnamefont {Lasseter}}, \bibinfo {author} {\bibfnamefont {D.}~\bibnamefont {Mallick}}, \bibinfo {author} {\bibfnamefont {J.}~\bibnamefont {Cook}}, \bibinfo {author} {\bibfnamefont {X.}~\bibnamefont {Yuan}}, \bibinfo {author} {\bibfnamefont {S.}~\bibnamefont {Oh}}, \bibinfo {author} {\bibfnamefont {G.}~\bibnamefont {Eres}}, \bibinfo {author} {\bibfnamefont {R.~G.}\ \bibnamefont {Moore}}, \bibinfo {author} {\bibfnamefont {T.}~\bibnamefont {Prokscha}}, \bibinfo {author} {\bibfnamefont {A.}~\bibnamefont {Suter}}, \bibinfo {author} {\bibfnamefont {S.}~\bibnamefont {Randolph}}, \bibinfo {author} {\bibfnamefont {J.~S.}\ \bibnamefont {Gardner}}, \bibinfo {author} {\bibfnamefont {P.~D.}\ \bibnamefont
  {Rack}}, \bibinfo {author} {\bibfnamefont {T.~Z.}\ \bibnamefont {Ward}}, \bibinfo {author} {\bibfnamefont {Z.}~\bibnamefont {Salman}}, \bibinfo {author} {\bibfnamefont {A.}~\bibnamefont {Georgescu}},\ and\ \bibinfo {author} {\bibfnamefont {M.}~\bibnamefont {Brahlek}},\ }\bibfield  {title} {\bibinfo {title} {Origins of ferromagnetism in helium-implanted {PdCoO$_2$}},\ }\href {https://doi.org/10.1103/wvk5-ht93} {\bibfield  {journal} {\bibinfo  {journal} {Physical Review Materials}\ }\textbf {\bibinfo {volume} {10}},\ \bibinfo {pages} {094405} (\bibinfo {year} {2026})}\BibitemShut {NoStop}%
\bibitem [{\citenamefont {Guo}\ \emph {et~al.}(2015)\citenamefont {Guo}, \citenamefont {Dong}, \citenamefont {Rack}, \citenamefont {Budai}, \citenamefont {Beekman}, \citenamefont {Gai}, \citenamefont {Siemons}, \citenamefont {Gonzalez}, \citenamefont {Timilsina}, \citenamefont {Wong}, \citenamefont {Herklotz}, \citenamefont {Snijders}, \citenamefont {Dagotto},\ and\ \citenamefont {Ward}}]{Guo2015}%
  \BibitemOpen
  \bibfield  {author} {\bibinfo {author} {\bibfnamefont {H.}~\bibnamefont {Guo}}, \bibinfo {author} {\bibfnamefont {S.}~\bibnamefont {Dong}}, \bibinfo {author} {\bibfnamefont {P.~D.}\ \bibnamefont {Rack}}, \bibinfo {author} {\bibfnamefont {J.~D.}\ \bibnamefont {Budai}}, \bibinfo {author} {\bibfnamefont {C.}~\bibnamefont {Beekman}}, \bibinfo {author} {\bibfnamefont {Z.}~\bibnamefont {Gai}}, \bibinfo {author} {\bibfnamefont {W.}~\bibnamefont {Siemons}}, \bibinfo {author} {\bibfnamefont {C.~M.}\ \bibnamefont {Gonzalez}}, \bibinfo {author} {\bibfnamefont {R.}~\bibnamefont {Timilsina}}, \bibinfo {author} {\bibfnamefont {A.~T.}\ \bibnamefont {Wong}}, \bibinfo {author} {\bibfnamefont {A.}~\bibnamefont {Herklotz}}, \bibinfo {author} {\bibfnamefont {P.~C.}\ \bibnamefont {Snijders}}, \bibinfo {author} {\bibfnamefont {E.}~\bibnamefont {Dagotto}},\ and\ \bibinfo {author} {\bibfnamefont {T.~Z.}\ \bibnamefont {Ward}},\ }\bibfield  {title} {\bibinfo {title} {Strain doping: Reversible single-axis control of a complex oxide
  lattice via helium implantation},\ }\href {https://doi.org/10.1103/PhysRevLett.114.256801} {\bibfield  {journal} {\bibinfo  {journal} {Physical Review Letters}\ }\textbf {\bibinfo {volume} {114}},\ \bibinfo {pages} {256801} (\bibinfo {year} {2015})}\BibitemShut {NoStop}%
\bibitem [{\citenamefont {Herklotz}\ \emph {et~al.}(2018)\citenamefont {Herklotz}, \citenamefont {Gai}, \citenamefont {Sharma}, \citenamefont {Huon}, \citenamefont {Rus}, \citenamefont {Sun}, \citenamefont {Shen}, \citenamefont {Rack},\ and\ \citenamefont {Ward}}]{Herklotz2018Anisotropy}%
  \BibitemOpen
  \bibfield  {author} {\bibinfo {author} {\bibfnamefont {A.}~\bibnamefont {Herklotz}}, \bibinfo {author} {\bibfnamefont {Z.}~\bibnamefont {Gai}}, \bibinfo {author} {\bibfnamefont {Y.}~\bibnamefont {Sharma}}, \bibinfo {author} {\bibfnamefont {A.}~\bibnamefont {Huon}}, \bibinfo {author} {\bibfnamefont {S.~F.}\ \bibnamefont {Rus}}, \bibinfo {author} {\bibfnamefont {L.}~\bibnamefont {Sun}}, \bibinfo {author} {\bibfnamefont {J.}~\bibnamefont {Shen}}, \bibinfo {author} {\bibfnamefont {P.~D.}\ \bibnamefont {Rack}},\ and\ \bibinfo {author} {\bibfnamefont {T.~Z.}\ \bibnamefont {Ward}},\ }\bibfield  {title} {\bibinfo {title} {Designing magnetic anisotropy through strain doping},\ }\href {https://doi.org/10.1002/advs.201800356} {\bibfield  {journal} {\bibinfo  {journal} {Advanced Science}\ }\textbf {\bibinfo {volume} {5}},\ \bibinfo {pages} {1800356} (\bibinfo {year} {2018})}\BibitemShut {NoStop}%
\bibitem [{\citenamefont {Skoropata}\ \emph {et~al.}(2021)\citenamefont {Skoropata}, \citenamefont {Mazza}, \citenamefont {Herklotz}, \citenamefont {Ok}, \citenamefont {Eres}, \citenamefont {Brahlek}, \citenamefont {Charlton}, \citenamefont {Lee},\ and\ \citenamefont {Ward}}]{Skoropata2021}%
  \BibitemOpen
  \bibfield  {author} {\bibinfo {author} {\bibfnamefont {E.}~\bibnamefont {Skoropata}}, \bibinfo {author} {\bibfnamefont {A.~R.}\ \bibnamefont {Mazza}}, \bibinfo {author} {\bibfnamefont {A.}~\bibnamefont {Herklotz}}, \bibinfo {author} {\bibfnamefont {J.~M.}\ \bibnamefont {Ok}}, \bibinfo {author} {\bibfnamefont {G.}~\bibnamefont {Eres}}, \bibinfo {author} {\bibfnamefont {M.}~\bibnamefont {Brahlek}}, \bibinfo {author} {\bibfnamefont {T.~R.}\ \bibnamefont {Charlton}}, \bibinfo {author} {\bibfnamefont {H.~N.}\ \bibnamefont {Lee}},\ and\ \bibinfo {author} {\bibfnamefont {T.~Z.}\ \bibnamefont {Ward}},\ }\bibfield  {title} {\bibinfo {title} {Post-synthesis control of {Berry} phase driven magnetotransport in {SrRuO$_3$} films},\ }\href {https://doi.org/10.1103/PhysRevB.103.085121} {\bibfield  {journal} {\bibinfo  {journal} {Physical Review B}\ }\textbf {\bibinfo {volume} {103}},\ \bibinfo {pages} {085121} (\bibinfo {year} {2021})}\BibitemShut {NoStop}%
\bibitem [{\citenamefont {Herklotz}\ \emph {et~al.}(2019)\citenamefont {Herklotz}, \citenamefont {Rus}, \citenamefont {Sohn}, \citenamefont {KC}, \citenamefont {Cooper}, \citenamefont {Guo},\ and\ \citenamefont {Ward}}]{Herklotz2019Optical}%
  \BibitemOpen
  \bibfield  {author} {\bibinfo {author} {\bibfnamefont {A.}~\bibnamefont {Herklotz}}, \bibinfo {author} {\bibfnamefont {S.~F.}\ \bibnamefont {Rus}}, \bibinfo {author} {\bibfnamefont {C.}~\bibnamefont {Sohn}}, \bibinfo {author} {\bibfnamefont {S.}~\bibnamefont {KC}}, \bibinfo {author} {\bibfnamefont {V.~R.}\ \bibnamefont {Cooper}}, \bibinfo {author} {\bibfnamefont {E.-J.}\ \bibnamefont {Guo}},\ and\ \bibinfo {author} {\bibfnamefont {T.~Z.}\ \bibnamefont {Ward}},\ }\bibfield  {title} {\bibinfo {title} {Optical response of {BiFeO$_3$} films subjected to uniaxial strain},\ }\href {https://doi.org/10.1103/PhysRevMaterials.3.094410} {\bibfield  {journal} {\bibinfo  {journal} {Physical Review Materials}\ }\textbf {\bibinfo {volume} {3}},\ \bibinfo {pages} {094410} (\bibinfo {year} {2019})}\BibitemShut {NoStop}%
\bibitem [{\citenamefont {Toulouse}\ \emph {et~al.}(2021)\citenamefont {Toulouse}, \citenamefont {Fischer}, \citenamefont {Farokhipoor}, \citenamefont {Yedra}, \citenamefont {Carl{\`a}}, \citenamefont {Jarnac}, \citenamefont {Elkaim}, \citenamefont {Fertey}, \citenamefont {Audinot}, \citenamefont {Wirtz}, \citenamefont {Noheda}, \citenamefont {Garcia}, \citenamefont {Fusil}, \citenamefont {Peral~Alonso}, \citenamefont {Guennou},\ and\ \citenamefont {Kreisel}}]{Toulouse2021}%
  \BibitemOpen
  \bibfield  {author} {\bibinfo {author} {\bibfnamefont {C.}~\bibnamefont {Toulouse}}, \bibinfo {author} {\bibfnamefont {J.}~\bibnamefont {Fischer}}, \bibinfo {author} {\bibfnamefont {S.}~\bibnamefont {Farokhipoor}}, \bibinfo {author} {\bibfnamefont {L.}~\bibnamefont {Yedra}}, \bibinfo {author} {\bibfnamefont {F.}~\bibnamefont {Carl{\`a}}}, \bibinfo {author} {\bibfnamefont {A.}~\bibnamefont {Jarnac}}, \bibinfo {author} {\bibfnamefont {E.}~\bibnamefont {Elkaim}}, \bibinfo {author} {\bibfnamefont {P.}~\bibnamefont {Fertey}}, \bibinfo {author} {\bibfnamefont {J.-N.}\ \bibnamefont {Audinot}}, \bibinfo {author} {\bibfnamefont {T.}~\bibnamefont {Wirtz}}, \bibinfo {author} {\bibfnamefont {B.}~\bibnamefont {Noheda}}, \bibinfo {author} {\bibfnamefont {V.}~\bibnamefont {Garcia}}, \bibinfo {author} {\bibfnamefont {S.}~\bibnamefont {Fusil}}, \bibinfo {author} {\bibfnamefont {I.}~\bibnamefont {Peral~Alonso}}, \bibinfo {author} {\bibfnamefont {M.}~\bibnamefont {Guennou}},\ and\ \bibinfo {author} {\bibfnamefont
  {J.}~\bibnamefont {Kreisel}},\ }\bibfield  {title} {\bibinfo {title} {Patterning enhanced tetragonality in {BiFeO$_3$} thin films with effective negative pressure by helium implantation},\ }\href {https://doi.org/10.1103/PhysRevMaterials.5.024404} {\bibfield  {journal} {\bibinfo  {journal} {Physical Review Materials}\ }\textbf {\bibinfo {volume} {5}},\ \bibinfo {pages} {024404} (\bibinfo {year} {2021})}\BibitemShut {NoStop}%
\bibitem [{\citenamefont {Wang}\ \emph {et~al.}(2019{\natexlab{a}})\citenamefont {Wang}, \citenamefont {Chang}, \citenamefont {Huang}, \citenamefont {Wang}, \citenamefont {Wu}, \citenamefont {Yang}, \citenamefont {Xu}, \citenamefont {Pandey}, \citenamefont {Zeng}, \citenamefont {B{\"o}ttger}, \citenamefont {Jeng}, \citenamefont {Zeng}, \citenamefont {Helm}, \citenamefont {Chu}, \citenamefont {Ganesh},\ and\ \citenamefont {Zhou}}]{Wang2019Disorder}%
  \BibitemOpen
  \bibfield  {author} {\bibinfo {author} {\bibfnamefont {C.}~\bibnamefont {Wang}}, \bibinfo {author} {\bibfnamefont {C.-H.}\ \bibnamefont {Chang}}, \bibinfo {author} {\bibfnamefont {A.}~\bibnamefont {Huang}}, \bibinfo {author} {\bibfnamefont {P.-C.}\ \bibnamefont {Wang}}, \bibinfo {author} {\bibfnamefont {P.-C.}\ \bibnamefont {Wu}}, \bibinfo {author} {\bibfnamefont {L.}~\bibnamefont {Yang}}, \bibinfo {author} {\bibfnamefont {C.}~\bibnamefont {Xu}}, \bibinfo {author} {\bibfnamefont {P.}~\bibnamefont {Pandey}}, \bibinfo {author} {\bibfnamefont {M.}~\bibnamefont {Zeng}}, \bibinfo {author} {\bibfnamefont {R.}~\bibnamefont {B{\"o}ttger}}, \bibinfo {author} {\bibfnamefont {H.-T.}\ \bibnamefont {Jeng}}, \bibinfo {author} {\bibfnamefont {Y.-J.}\ \bibnamefont {Zeng}}, \bibinfo {author} {\bibfnamefont {M.}~\bibnamefont {Helm}}, \bibinfo {author} {\bibfnamefont {Y.-H.}\ \bibnamefont {Chu}}, \bibinfo {author} {\bibfnamefont {R.}~\bibnamefont {Ganesh}},\ and\ \bibinfo {author} {\bibfnamefont {S.}~\bibnamefont {Zhou}},\
  }\bibfield  {title} {\bibinfo {title} {Tunable disorder and localization in the rare-earth nickelates},\ }\href {https://doi.org/10.1103/PhysRevMaterials.3.053801} {\bibfield  {journal} {\bibinfo  {journal} {Physical Review Materials}\ }\textbf {\bibinfo {volume} {3}},\ \bibinfo {pages} {053801} (\bibinfo {year} {2019}{\natexlab{a}})}\BibitemShut {NoStop}%
\bibitem [{\citenamefont {Gurrola}\ \emph {et~al.}(2024)\citenamefont {Gurrola}, \citenamefont {Cain}, \citenamefont {Oh}, \citenamefont {Brown}, \citenamefont {Jardali}, \citenamefont {Schoell}, \citenamefont {Yadav}, \citenamefont {Dong}, \citenamefont {Smyth}, \citenamefont {Pharr}, \citenamefont {Kumar}, \citenamefont {Xie}, \citenamefont {Hattar}, \citenamefont {Talin}, \citenamefont {Lu},\ and\ \citenamefont {Shamberger}}]{Gurrola2024}%
  \BibitemOpen
  \bibfield  {author} {\bibinfo {author} {\bibfnamefont {R.~M.}\ \bibnamefont {Gurrola}}, \bibinfo {author} {\bibfnamefont {J.~M.}\ \bibnamefont {Cain}}, \bibinfo {author} {\bibfnamefont {S.}~\bibnamefont {Oh}}, \bibinfo {author} {\bibfnamefont {T.~D.}\ \bibnamefont {Brown}}, \bibinfo {author} {\bibfnamefont {F.}~\bibnamefont {Jardali}}, \bibinfo {author} {\bibfnamefont {R.~M.}\ \bibnamefont {Schoell}}, \bibinfo {author} {\bibfnamefont {D.~R.}\ \bibnamefont {Yadav}}, \bibinfo {author} {\bibfnamefont {J.}~\bibnamefont {Dong}}, \bibinfo {author} {\bibfnamefont {C.~M.}\ \bibnamefont {Smyth}}, \bibinfo {author} {\bibfnamefont {M.}~\bibnamefont {Pharr}}, \bibinfo {author} {\bibfnamefont {S.}~\bibnamefont {Kumar}}, \bibinfo {author} {\bibfnamefont {K.}~\bibnamefont {Xie}}, \bibinfo {author} {\bibfnamefont {K.}~\bibnamefont {Hattar}}, \bibinfo {author} {\bibfnamefont {A.~A.}\ \bibnamefont {Talin}}, \bibinfo {author} {\bibfnamefont {T.-M.}\ \bibnamefont {Lu}},\ and\ \bibinfo {author} {\bibfnamefont {P.~J.}\
  \bibnamefont {Shamberger}},\ }\bibfield  {title} {\bibinfo {title} {Selective modulation of electronic transport in {VO$_2$} induced by 10 kev helium ion irradiation},\ }\href {https://doi.org/10.1063/5.0189562} {\bibfield  {journal} {\bibinfo  {journal} {Journal of Applied Physics}\ }\textbf {\bibinfo {volume} {135}},\ \bibinfo {pages} {125109} (\bibinfo {year} {2024})}\BibitemShut {NoStop}%
\bibitem [{\citenamefont {Georgescu}\ \emph {et~al.}(2019)\citenamefont {Georgescu}, \citenamefont {Peil}, \citenamefont {Disa}, \citenamefont {Georges},\ and\ \citenamefont {Millis}}]{Georgescu2019}%
  \BibitemOpen
  \bibfield  {author} {\bibinfo {author} {\bibfnamefont {A.~B.}\ \bibnamefont {Georgescu}}, \bibinfo {author} {\bibfnamefont {O.~E.}\ \bibnamefont {Peil}}, \bibinfo {author} {\bibfnamefont {A.~S.}\ \bibnamefont {Disa}}, \bibinfo {author} {\bibfnamefont {A.}~\bibnamefont {Georges}},\ and\ \bibinfo {author} {\bibfnamefont {A.~J.}\ \bibnamefont {Millis}},\ }\bibfield  {title} {\bibinfo {title} {Disentangling lattice and electronic contributions to the metal--insulator transition from bulk vs. layer confined {RNiO$_3$}},\ }\href {https://doi.org/10.1073/pnas.1818728116} {\bibfield  {journal} {\bibinfo  {journal} {Proceedings of the National Academy of Sciences}\ }\textbf {\bibinfo {volume} {116}},\ \bibinfo {pages} {14434} (\bibinfo {year} {2019})}\BibitemShut {NoStop}%
\bibitem [{\citenamefont {Spring}\ \emph {et~al.}(2025)\citenamefont {Spring}, \citenamefont {Fedorova}, \citenamefont {Georgescu}, \citenamefont {Vogel}, \citenamefont {De~Luca}, \citenamefont {J{\"o}hr}, \citenamefont {Piamonteze}, \citenamefont {Rossell}, \citenamefont {{\'I}{\~n}iguez-Gonz{\'a}lez},\ and\ \citenamefont {Gibert}}]{Spring2025}%
  \BibitemOpen
  \bibfield  {author} {\bibinfo {author} {\bibfnamefont {J.}~\bibnamefont {Spring}}, \bibinfo {author} {\bibfnamefont {N.}~\bibnamefont {Fedorova}}, \bibinfo {author} {\bibfnamefont {A.~B.}\ \bibnamefont {Georgescu}}, \bibinfo {author} {\bibfnamefont {A.}~\bibnamefont {Vogel}}, \bibinfo {author} {\bibfnamefont {G.}~\bibnamefont {De~Luca}}, \bibinfo {author} {\bibfnamefont {S.}~\bibnamefont {J{\"o}hr}}, \bibinfo {author} {\bibfnamefont {C.}~\bibnamefont {Piamonteze}}, \bibinfo {author} {\bibfnamefont {M.~D.}\ \bibnamefont {Rossell}}, \bibinfo {author} {\bibfnamefont {J.}~\bibnamefont {{\'I}{\~n}iguez-Gonz{\'a}lez}},\ and\ \bibinfo {author} {\bibfnamefont {M.}~\bibnamefont {Gibert}},\ }\bibfield  {title} {\bibinfo {title} {Engineering the magnetic transition temperatures and the rare earth exchange interaction in oxide heterostructures},\ }\href {https://doi.org/10.1021/acsnano.4c07252} {\bibfield  {journal} {\bibinfo  {journal} {ACS Nano}\ }\textbf {\bibinfo {volume} {19}},\ \bibinfo {pages} {14652} (\bibinfo
  {year} {2025})}\BibitemShut {NoStop}%
\bibitem [{\citenamefont {Kang}\ \emph {et~al.}(2022)\citenamefont {Kang}, \citenamefont {Jang}, \citenamefont {Morozovska}, \citenamefont {Kwon}, \citenamefont {Jin}, \citenamefont {Kim}, \citenamefont {Bae}, \citenamefont {Wang}, \citenamefont {Yang}, \citenamefont {Belianinov}, \citenamefont {Randolph}, \citenamefont {Eliseev}, \citenamefont {Collins}, \citenamefont {Park}, \citenamefont {Jo}, \citenamefont {Jung}, \citenamefont {Go}, \citenamefont {Cho}, \citenamefont {Choi}, \citenamefont {Jang}, \citenamefont {Kim}, \citenamefont {Jeong}, \citenamefont {Lee}, \citenamefont {Ovchinnikova}, \citenamefont {Heo}, \citenamefont {Kalinin}, \citenamefont {Kim},\ and\ \citenamefont {Kim}}]{Kang2022}%
  \BibitemOpen
  \bibfield  {author} {\bibinfo {author} {\bibfnamefont {S.}~\bibnamefont {Kang}}, \bibinfo {author} {\bibfnamefont {W.-S.}\ \bibnamefont {Jang}}, \bibinfo {author} {\bibfnamefont {A.~N.}\ \bibnamefont {Morozovska}}, \bibinfo {author} {\bibfnamefont {O.}~\bibnamefont {Kwon}}, \bibinfo {author} {\bibfnamefont {Y.}~\bibnamefont {Jin}}, \bibinfo {author} {\bibfnamefont {Y.-H.}\ \bibnamefont {Kim}}, \bibinfo {author} {\bibfnamefont {H.}~\bibnamefont {Bae}}, \bibinfo {author} {\bibfnamefont {C.}~\bibnamefont {Wang}}, \bibinfo {author} {\bibfnamefont {S.-H.}\ \bibnamefont {Yang}}, \bibinfo {author} {\bibfnamefont {A.}~\bibnamefont {Belianinov}}, \bibinfo {author} {\bibfnamefont {S.}~\bibnamefont {Randolph}}, \bibinfo {author} {\bibfnamefont {E.~A.}\ \bibnamefont {Eliseev}}, \bibinfo {author} {\bibfnamefont {L.}~\bibnamefont {Collins}}, \bibinfo {author} {\bibfnamefont {Y.}~\bibnamefont {Park}}, \bibinfo {author} {\bibfnamefont {S.}~\bibnamefont {Jo}}, \bibinfo {author} {\bibfnamefont {M.-H.}\ \bibnamefont {Jung}},
  \bibinfo {author} {\bibfnamefont {K.-J.}\ \bibnamefont {Go}}, \bibinfo {author} {\bibfnamefont {H.~W.}\ \bibnamefont {Cho}}, \bibinfo {author} {\bibfnamefont {S.-Y.}\ \bibnamefont {Choi}}, \bibinfo {author} {\bibfnamefont {J.~H.}\ \bibnamefont {Jang}}, \bibinfo {author} {\bibfnamefont {S.}~\bibnamefont {Kim}}, \bibinfo {author} {\bibfnamefont {H.~Y.}\ \bibnamefont {Jeong}}, \bibinfo {author} {\bibfnamefont {J.}~\bibnamefont {Lee}}, \bibinfo {author} {\bibfnamefont {O.~S.}\ \bibnamefont {Ovchinnikova}}, \bibinfo {author} {\bibfnamefont {J.}~\bibnamefont {Heo}}, \bibinfo {author} {\bibfnamefont {S.~V.}\ \bibnamefont {Kalinin}}, \bibinfo {author} {\bibfnamefont {Y.-M.}\ \bibnamefont {Kim}},\ and\ \bibinfo {author} {\bibfnamefont {Y.}~\bibnamefont {Kim}},\ }\bibfield  {title} {\bibinfo {title} {Highly enhanced ferroelectricity in {HfO$_2$}-based ferroelectric thin film by light ion bombardment},\ }\href {https://doi.org/10.1126/science.abk3195} {\bibfield  {journal} {\bibinfo  {journal} {Science}\ }\textbf
  {\bibinfo {volume} {376}},\ \bibinfo {pages} {731} (\bibinfo {year} {2022})}\BibitemShut {NoStop}%
\bibitem [{\citenamefont {Wang}\ \emph {et~al.}(2020)\citenamefont {Wang}, \citenamefont {Chang}, \citenamefont {Herklotz}, \citenamefont {Chen}, \citenamefont {Ganss}, \citenamefont {Kentsch}, \citenamefont {Chen}, \citenamefont {Gao}, \citenamefont {Zeng}, \citenamefont {Hellwig}, \citenamefont {Helm}, \citenamefont {Gemming}, \citenamefont {Chu},\ and\ \citenamefont {Zhou}}]{Wang2020Hall}%
  \BibitemOpen
  \bibfield  {author} {\bibinfo {author} {\bibfnamefont {C.}~\bibnamefont {Wang}}, \bibinfo {author} {\bibfnamefont {C.-H.}\ \bibnamefont {Chang}}, \bibinfo {author} {\bibfnamefont {A.}~\bibnamefont {Herklotz}}, \bibinfo {author} {\bibfnamefont {C.}~\bibnamefont {Chen}}, \bibinfo {author} {\bibfnamefont {F.}~\bibnamefont {Ganss}}, \bibinfo {author} {\bibfnamefont {U.}~\bibnamefont {Kentsch}}, \bibinfo {author} {\bibfnamefont {D.}~\bibnamefont {Chen}}, \bibinfo {author} {\bibfnamefont {X.}~\bibnamefont {Gao}}, \bibinfo {author} {\bibfnamefont {Y.-J.}\ \bibnamefont {Zeng}}, \bibinfo {author} {\bibfnamefont {O.}~\bibnamefont {Hellwig}}, \bibinfo {author} {\bibfnamefont {M.}~\bibnamefont {Helm}}, \bibinfo {author} {\bibfnamefont {S.}~\bibnamefont {Gemming}}, \bibinfo {author} {\bibfnamefont {Y.-H.}\ \bibnamefont {Chu}},\ and\ \bibinfo {author} {\bibfnamefont {S.}~\bibnamefont {Zhou}},\ }\bibfield  {title} {\bibinfo {title} {Topological hall effect in single thick {SrRuO$_3$} layers induced by defect
  engineering},\ }\href {https://doi.org/10.1002/aelm.202000184} {\bibfield  {journal} {\bibinfo  {journal} {Advanced Electronic Materials}\ }\textbf {\bibinfo {volume} {6}},\ \bibinfo {pages} {2000184} (\bibinfo {year} {2020})}\BibitemShut {NoStop}%
\bibitem [{\citenamefont {Bhattacharya}\ and\ \citenamefont {May}(2014)}]{Bhattacharya2014}%
  \BibitemOpen
  \bibfield  {author} {\bibinfo {author} {\bibfnamefont {A.}~\bibnamefont {Bhattacharya}}\ and\ \bibinfo {author} {\bibfnamefont {S.~J.}\ \bibnamefont {May}},\ }\bibfield  {title} {\bibinfo {title} {Magnetic oxide heterostructures},\ }\href {https://doi.org/https://doi.org/10.1146/annurev-matsci-070813-113447} {\bibfield  {journal} {\bibinfo  {journal} {Annual Review of Materials Research}\ }\textbf {\bibinfo {volume} {44}},\ \bibinfo {pages} {65} (\bibinfo {year} {2014})}\BibitemShut {NoStop}%
\bibitem [{\citenamefont {Wang}\ \emph {et~al.}(2019{\natexlab{b}})\citenamefont {Wang}, \citenamefont {Zhang}, \citenamefont {Deepak}, \citenamefont {Chen}, \citenamefont {Fouchet}, \citenamefont {Duan}, \citenamefont {Hilliard}, \citenamefont {Kentsch}, \citenamefont {Chen}, \citenamefont {Zeng}, \citenamefont {Gao}, \citenamefont {Zeng}, \citenamefont {Helm}, \citenamefont {Prellier},\ and\ \citenamefont {Zhou}}]{Wang2019}%
  \BibitemOpen
  \bibfield  {author} {\bibinfo {author} {\bibfnamefont {C.}~\bibnamefont {Wang}}, \bibinfo {author} {\bibfnamefont {H.}~\bibnamefont {Zhang}}, \bibinfo {author} {\bibfnamefont {K.}~\bibnamefont {Deepak}}, \bibinfo {author} {\bibfnamefont {C.}~\bibnamefont {Chen}}, \bibinfo {author} {\bibfnamefont {A.}~\bibnamefont {Fouchet}}, \bibinfo {author} {\bibfnamefont {J.}~\bibnamefont {Duan}}, \bibinfo {author} {\bibfnamefont {D.}~\bibnamefont {Hilliard}}, \bibinfo {author} {\bibfnamefont {U.}~\bibnamefont {Kentsch}}, \bibinfo {author} {\bibfnamefont {D.}~\bibnamefont {Chen}}, \bibinfo {author} {\bibfnamefont {M.}~\bibnamefont {Zeng}}, \bibinfo {author} {\bibfnamefont {X.}~\bibnamefont {Gao}}, \bibinfo {author} {\bibfnamefont {Y.-J.}\ \bibnamefont {Zeng}}, \bibinfo {author} {\bibfnamefont {M.}~\bibnamefont {Helm}}, \bibinfo {author} {\bibfnamefont {W.}~\bibnamefont {Prellier}},\ and\ \bibinfo {author} {\bibfnamefont {S.}~\bibnamefont {Zhou}},\ }\bibfield  {title} {\bibinfo {title} {Tuning the metal-insulator
  transition in epitaxial {SrVO$_3$} films by uniaxial strain},\ }\href {https://doi.org/10.1103/PhysRevMaterials.3.115001} {\bibfield  {journal} {\bibinfo  {journal} {Physical Review Materials}\ }\textbf {\bibinfo {volume} {3}},\ \bibinfo {pages} {115001} (\bibinfo {year} {2019}{\natexlab{b}})}\BibitemShut {NoStop}%
\bibitem [{\citenamefont {Pickem}\ \emph {et~al.}(2026)\citenamefont {Pickem}, \citenamefont {Held},\ and\ \citenamefont {Tomczak}}]{pickem2026}%
  \BibitemOpen
  \bibfield  {author} {\bibinfo {author} {\bibfnamefont {M.}~\bibnamefont {Pickem}}, \bibinfo {author} {\bibfnamefont {K.}~\bibnamefont {Held}},\ and\ \bibinfo {author} {\bibfnamefont {J.~M.}\ \bibnamefont {Tomczak}},\ }\href {https://arxiv.org/abs/2608.14362} {\bibinfo {title} {Tuning crystal-fields by {He}-irradiation and orbital {Widom} line in {SrVO$_3$} films}} (\bibinfo {year} {2026}),\ \Eprint {https://arxiv.org/abs/2608.14362} {arXiv:2608.14362} \BibitemShut {NoStop}%
\bibitem [{\citenamefont {Giannozzi}\ \emph {et~al.}(2017)\citenamefont {Giannozzi}, \citenamefont {Andreussi}, \citenamefont {Brumme}, \citenamefont {Bunau}, \citenamefont {Buongiorno~Nardelli}, \citenamefont {Calandra}, \citenamefont {Car}, \citenamefont {Cavazzoni}, \citenamefont {Ceresoli}, \citenamefont {Cococcioni}, \citenamefont {Colonna}, \citenamefont {Carnimeo}, \citenamefont {Dal~Corso}, \citenamefont {de~Gironcoli}, \citenamefont {Delugas}, \citenamefont {DiStasio}, \citenamefont {Ferretti}, \citenamefont {Floris}, \citenamefont {Fratesi}, \citenamefont {Fugallo}, \citenamefont {Gebauer}, \citenamefont {Gerstmann}, \citenamefont {Giustino}, \citenamefont {Gorni}, \citenamefont {Jia}, \citenamefont {Kawamura}, \citenamefont {Ko}, \citenamefont {Kokalj}, \citenamefont {K{\"u}{\c{c}}{\"u}kbenli}, \citenamefont {Lazzeri}, \citenamefont {Marsili}, \citenamefont {Marzari}, \citenamefont {Mauri}, \citenamefont {Nguyen}, \citenamefont {Nguyen}, \citenamefont {Otero-de-la Roza}, \citenamefont {Paulatto},
  \citenamefont {Ponc{\'e}}, \citenamefont {Rocca}, \citenamefont {Sabatini}, \citenamefont {Santra}, \citenamefont {Schlipf}, \citenamefont {Seitsonen}, \citenamefont {Smogunov}, \citenamefont {Timrov}, \citenamefont {Thonhauser}, \citenamefont {Umari}, \citenamefont {Vast}, \citenamefont {Wu},\ and\ \citenamefont {Baroni}}]{giannozzi2017}%
  \BibitemOpen
  \bibfield  {author} {\bibinfo {author} {\bibfnamefont {P.}~\bibnamefont {Giannozzi}}, \bibinfo {author} {\bibfnamefont {O.}~\bibnamefont {Andreussi}}, \bibinfo {author} {\bibfnamefont {T.}~\bibnamefont {Brumme}}, \bibinfo {author} {\bibfnamefont {O.}~\bibnamefont {Bunau}}, \bibinfo {author} {\bibfnamefont {M.}~\bibnamefont {Buongiorno~Nardelli}}, \bibinfo {author} {\bibfnamefont {M.}~\bibnamefont {Calandra}}, \bibinfo {author} {\bibfnamefont {R.}~\bibnamefont {Car}}, \bibinfo {author} {\bibfnamefont {C.}~\bibnamefont {Cavazzoni}}, \bibinfo {author} {\bibfnamefont {D.}~\bibnamefont {Ceresoli}}, \bibinfo {author} {\bibfnamefont {M.}~\bibnamefont {Cococcioni}}, \bibinfo {author} {\bibfnamefont {N.}~\bibnamefont {Colonna}}, \bibinfo {author} {\bibfnamefont {I.}~\bibnamefont {Carnimeo}}, \bibinfo {author} {\bibfnamefont {A.}~\bibnamefont {Dal~Corso}}, \bibinfo {author} {\bibfnamefont {S.}~\bibnamefont {de~Gironcoli}}, \bibinfo {author} {\bibfnamefont {P.}~\bibnamefont {Delugas}}, \bibinfo {author} {\bibfnamefont
  {R.~A.}\ \bibnamefont {DiStasio}, \bibfnamefont {Jr.}}, \bibinfo {author} {\bibfnamefont {A.}~\bibnamefont {Ferretti}}, \bibinfo {author} {\bibfnamefont {A.}~\bibnamefont {Floris}}, \bibinfo {author} {\bibfnamefont {G.}~\bibnamefont {Fratesi}}, \bibinfo {author} {\bibfnamefont {G.}~\bibnamefont {Fugallo}}, \bibinfo {author} {\bibfnamefont {R.}~\bibnamefont {Gebauer}}, \bibinfo {author} {\bibfnamefont {U.}~\bibnamefont {Gerstmann}}, \bibinfo {author} {\bibfnamefont {F.}~\bibnamefont {Giustino}}, \bibinfo {author} {\bibfnamefont {T.}~\bibnamefont {Gorni}}, \bibinfo {author} {\bibfnamefont {J.}~\bibnamefont {Jia}}, \bibinfo {author} {\bibfnamefont {M.}~\bibnamefont {Kawamura}}, \bibinfo {author} {\bibfnamefont {H.-Y.}\ \bibnamefont {Ko}}, \bibinfo {author} {\bibfnamefont {A.}~\bibnamefont {Kokalj}}, \bibinfo {author} {\bibfnamefont {E.}~\bibnamefont {K{\"u}{\c{c}}{\"u}kbenli}}, \bibinfo {author} {\bibfnamefont {M.}~\bibnamefont {Lazzeri}}, \bibinfo {author} {\bibfnamefont {M.}~\bibnamefont {Marsili}}, \bibinfo
  {author} {\bibfnamefont {N.}~\bibnamefont {Marzari}}, \bibinfo {author} {\bibfnamefont {F.}~\bibnamefont {Mauri}}, \bibinfo {author} {\bibfnamefont {N.~L.}\ \bibnamefont {Nguyen}}, \bibinfo {author} {\bibfnamefont {H.-V.}\ \bibnamefont {Nguyen}}, \bibinfo {author} {\bibfnamefont {A.}~\bibnamefont {Otero-de-la Roza}}, \bibinfo {author} {\bibfnamefont {L.}~\bibnamefont {Paulatto}}, \bibinfo {author} {\bibfnamefont {S.}~\bibnamefont {Ponc{\'e}}}, \bibinfo {author} {\bibfnamefont {D.}~\bibnamefont {Rocca}}, \bibinfo {author} {\bibfnamefont {R.}~\bibnamefont {Sabatini}}, \bibinfo {author} {\bibfnamefont {B.}~\bibnamefont {Santra}}, \bibinfo {author} {\bibfnamefont {M.}~\bibnamefont {Schlipf}}, \bibinfo {author} {\bibfnamefont {A.~P.}\ \bibnamefont {Seitsonen}}, \bibinfo {author} {\bibfnamefont {A.}~\bibnamefont {Smogunov}}, \bibinfo {author} {\bibfnamefont {I.}~\bibnamefont {Timrov}}, \bibinfo {author} {\bibfnamefont {T.}~\bibnamefont {Thonhauser}}, \bibinfo {author} {\bibfnamefont {P.}~\bibnamefont {Umari}},
  \bibinfo {author} {\bibfnamefont {N.}~\bibnamefont {Vast}}, \bibinfo {author} {\bibfnamefont {X.}~\bibnamefont {Wu}},\ and\ \bibinfo {author} {\bibfnamefont {S.}~\bibnamefont {Baroni}},\ }\bibfield  {title} {\bibinfo {title} {Advanced capabilities for materials modelling with {Quantum ESPRESSO}},\ }\href {https://doi.org/10.1088/1361-648X/aa8f79} {\bibfield  {journal} {\bibinfo  {journal} {Journal of Physics: Condensed Matter}\ }\textbf {\bibinfo {volume} {29}},\ \bibinfo {pages} {465901} (\bibinfo {year} {2017})}\BibitemShut {NoStop}%
\bibitem [{\citenamefont {Perdew}\ \emph {et~al.}(1996)\citenamefont {Perdew}, \citenamefont {Burke},\ and\ \citenamefont {Ernzerhof}}]{Perdew1996}%
  \BibitemOpen
  \bibfield  {author} {\bibinfo {author} {\bibfnamefont {J.~P.}\ \bibnamefont {Perdew}}, \bibinfo {author} {\bibfnamefont {K.}~\bibnamefont {Burke}},\ and\ \bibinfo {author} {\bibfnamefont {M.}~\bibnamefont {Ernzerhof}},\ }\bibfield  {title} {\bibinfo {title} {{Generalized Gradient Approximation Made Simple}},\ }\href {https://doi.org/10.1103/PhysRevLett.77.3865} {\bibfield  {journal} {\bibinfo  {journal} {Physical Review Letters}\ }\textbf {\bibinfo {volume} {77}},\ \bibinfo {pages} {3865} (\bibinfo {year} {1996})}\BibitemShut {NoStop}%
\bibitem [{\citenamefont {Vanderbilt}(1990)}]{Vanderbilt1990}%
  \BibitemOpen
  \bibfield  {author} {\bibinfo {author} {\bibfnamefont {D.}~\bibnamefont {Vanderbilt}},\ }\bibfield  {title} {\bibinfo {title} {{Soft self-consistent pseudopotentials in a generalized eigenvalue formalism}},\ }\href {https://doi.org/10.1103/PhysRevB.41.7892} {\bibfield  {journal} {\bibinfo  {journal} {Physical Review B}\ }\textbf {\bibinfo {volume} {41}},\ \bibinfo {pages} {7892} (\bibinfo {year} {1990})}\BibitemShut {NoStop}%
\bibitem [{\citenamefont {Dudarev}\ \emph {et~al.}(1998)\citenamefont {Dudarev}, \citenamefont {Botton}, \citenamefont {Savrasov}, \citenamefont {Humphreys},\ and\ \citenamefont {Sutton}}]{Dudarev1998}%
  \BibitemOpen
  \bibfield  {author} {\bibinfo {author} {\bibfnamefont {S.~L.}\ \bibnamefont {Dudarev}}, \bibinfo {author} {\bibfnamefont {G.~A.}\ \bibnamefont {Botton}}, \bibinfo {author} {\bibfnamefont {S.~Y.}\ \bibnamefont {Savrasov}}, \bibinfo {author} {\bibfnamefont {C.~J.}\ \bibnamefont {Humphreys}},\ and\ \bibinfo {author} {\bibfnamefont {A.~P.}\ \bibnamefont {Sutton}},\ }\bibfield  {title} {\bibinfo {title} {{Electron-energy-loss spectra and the structural stability of nickel oxide: An LSDA+U study}},\ }\href {https://doi.org/10.1103/PhysRevB.57.1505} {\bibfield  {journal} {\bibinfo  {journal} {Physical Review B}\ }\textbf {\bibinfo {volume} {57}},\ \bibinfo {pages} {1505} (\bibinfo {year} {1998})}\BibitemShut {NoStop}%
\bibitem [{\citenamefont {Marzari}\ \emph {et~al.}(1999)\citenamefont {Marzari}, \citenamefont {Vanderbilt}, \citenamefont {De~Vita},\ and\ \citenamefont {Payne}}]{Marzari1999}%
  \BibitemOpen
  \bibfield  {author} {\bibinfo {author} {\bibfnamefont {N.}~\bibnamefont {Marzari}}, \bibinfo {author} {\bibfnamefont {D.}~\bibnamefont {Vanderbilt}}, \bibinfo {author} {\bibfnamefont {A.}~\bibnamefont {De~Vita}},\ and\ \bibinfo {author} {\bibfnamefont {M.~C.}\ \bibnamefont {Payne}},\ }\bibfield  {title} {\bibinfo {title} {Thermal contraction and disordering of the {Al(110)} surface},\ }\href {https://doi.org/10.1103/PhysRevLett.82.3296} {\bibfield  {journal} {\bibinfo  {journal} {Physical Review Letters}\ }\textbf {\bibinfo {volume} {82}},\ \bibinfo {pages} {3296} (\bibinfo {year} {1999})}\BibitemShut {NoStop}%
\end{thebibliography}%
\end{document}